\documentclass[journal]{IEEEtran}
\usepackage{etoolbox}
\usepackage{xcolor}
\usepackage{cite}
\usepackage[pdftex]{graphicx}
\usepackage[font=small,skip=2pt,justification=centering]{caption}
\DeclareGraphicsExtensions{.pdf,.jpeg,.png}

\usepackage{amsmath}
\usepackage{algorithm}
\usepackage[noend]{algpseudocode}
\usepackage{dsfont}
\usepackage{amsthm}
\usepackage{amssymb}
\usepackage{color}
\usepackage{array}
\usepackage{multirow}
\usepackage[caption=false,font=footnotesize]{subfig}
\newcommand\cmb[1]{\textcolor{black}{#1}}

\newcommand\boldx{\boldsymbol{x}}

\usepackage[compact]{titlesec}

\makeatletter
\patchcmd{\@maketitle}
  {\addvspace{0.5\baselineskip}\egroup}
  {\addvspace{-1\baselineskip}\egroup}
\makeatother

\begin{document}

\title{A Model-Agnostic Method for PMU Data Recovery Using Optimal Singular Value Thresholding}
\author{Shuchismita Biswas,~\IEEEmembership{Student Member,~IEEE,}
        and~Virgilo A. Centeno,~\IEEEmembership{Senior Member,~IEEE}
}

\markboth{Journal of \LaTeX\ Class Files,~Vol.~XX, No.~X, XXXX~XXXX}%
{Biswas \MakeLowercase{\textit{et al.}}: Model-Agnostic PMU Data Recovery Using Optimal Singular Value Thresholding}
\maketitle

\begin{abstract}
This paper presents a fast model-agnostic method for recovering noisy Phasor Measurement Unit (PMU) data streams with missing entries. The measurements are first transformed into a Page matrix, and the original signals are reconstructed using low-rank matrix estimation based on optimal singular value thresholding. Two variations of the recovery algorithm are shown- a) an offline block-processing method for imputing past measurements, and b) an online method for predicting future measurements. Information within a PMU channel (temporal correlation) as well as from different PMU channels in a network (spatial correlation) are utilized to recover degraded data.
The proposed method is fast, and needs no explicit knowledge of the underlying system model or measurement noise distribution. Performance of the recovery algorithms is illustrated using simulated measurements from the IEEE 39-bus test system as well as real measurements from an anonymized U.S. electric utility. Extensive numeric tests show that the original signals can be accurately recovered in the presence of additive noise, consecutive data drop as well as simultaneous data erasures across multiple PMU channels. 
\end{abstract}



\section{Introduction} \label{sec:1_intro}

\begin{figure}
    \centering
    \includegraphics[width=\columnwidth]{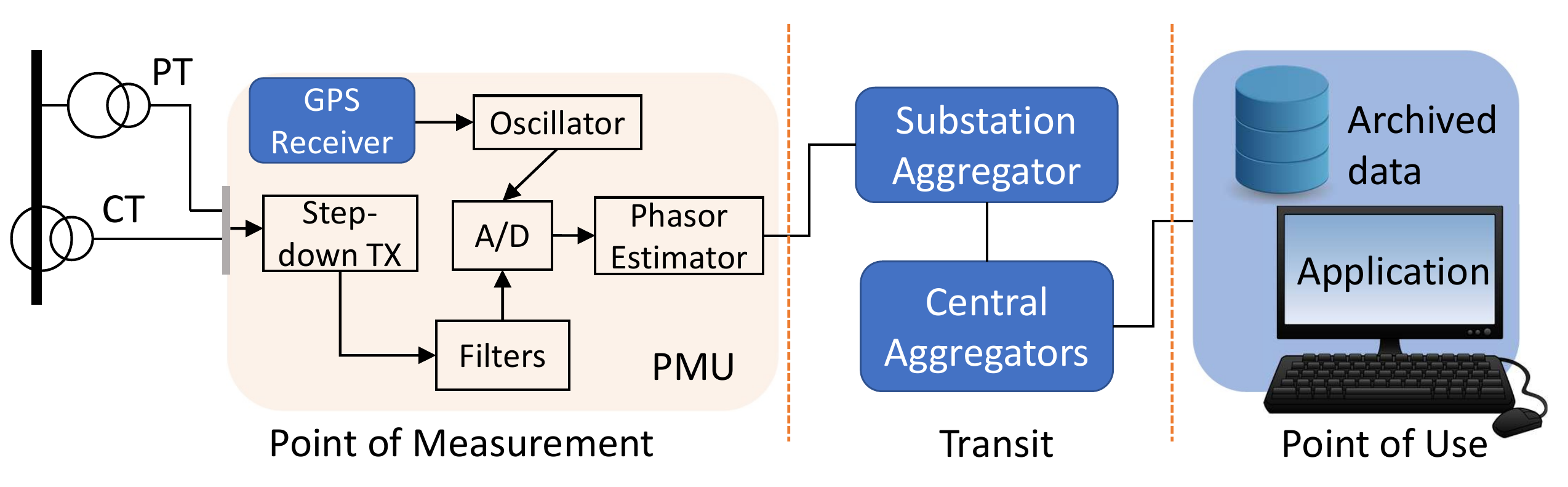}
    \caption{PMU data flow from point-of-measurement to point-of-use}
    \label{fig:pmu}
    \vspace{-0.25in}
\end{figure}

Phasor Measurement Units (PMUs) allow high-resolution insight into power systems dynamics through precise time-synchronized measurements \cite{PMU_OG}. In recent years, electric utilities have made great strides towards deploying PMUs in their networks and utilizing the reported measurements for wide-area situational awareness \cite{PMU_US}. Phasor measurements are used in both online (state estimation, remedial action schemes) and offline (model validation, contingency analysis, post-event diagnosis) applications \cite{onlinePMU},{\cite{jones2013synchrophasor}}, \cite{model_validation}. As shown in fig. \ref{fig:pmu}, to reach the point-of-use from the point-of-measurement, PMU data flows through various communication channels and intermediate data concentrators. Hence, the data is susceptible to channel congestion or component malfunction issues which lead to degraded information quality \cite{NASPI_data_quality}. Reliable measurements are critical to developing robust control and monitoring algorithms for the power grid, making fast and accurate data recovery critical as well.

Traditional model-based recovery methods are sensitive to the underlying model accuracy, and may need the knowledge of real-time network topology. \cmb{ Oftentimes, such methods assume the knowledge of transmission line parameters, and are adversely affected by inaccurate information \cite{Shi_lineparameters}. Another class of model-based methods use linear state estimation (LSE) for PMU data conditioning \cite{jones2013synchrophasor,LSE_WECC}. LSE-based data recovery is contingent on breaker status information and any error or latency in reporting may deteriorate performance \cite{WECC_thorp}.}  On the other hand, measurements from PMUs within a network exhibit spatial and temporal correlation that can be utilized to estimate missing samples without explicit knowledge of the power system model. Several recent papers have addressed model-agnostic PMU data recovery \cite{ tensor, tensor2, kalman, GEIRI, svd, PSCC_algo, MWang_Hankel}. Some of the proposed methods target imputation of data blocks \cite{GEIRI,tensor,tensor2}, while others are aimed at step-ahead prediction \cite{kalman, svd, PSCC_algo, MWang_Hankel}. Further, these recovery approaches broadly employ strategies based on - a) filtering \cite{kalman},  b) low rank matrix completion \cite{PSCC_algo,svd,GEIRI,MWang_Hankel}, or c) low rank tensor completion \cite{tensor,tensor2}.

In \cite{kalman}, the authors propose a kalman filter-based missing data estimation algorithm that predicts the value of an incoming sample using the last three measurements. The accuracy of this method deteriorates when more than three consecutive entries are missing. Moreover, as the algorithm processes one PMU channel at a time, information from other channels or nearby PMUs cannot be leveraged to reconstruct segments of consecutive missing data.
Matrix (tensor) estimation methods propose to  stack correlated PMU measurement channels together to construct matrices (tensors), whose low-rank property can be exploited to recover corrupt data. Tensor estimation methods are more computationally expensive, limiting their potential for real-time use. Methods to speed up tensor estimation have been explored in \cite{tensor,tensor2}. 

In the matrix estimation (ME) area, various strategies have been put forth for stacking measurement channels. In \cite{GEIRI}, the channels are stacked rowwise, and an iterative alternating direction method of multipliers is used to fill missing measurements. In \cite{svd}, the channels are stacked columnwise, and the spatial correlation among PMUs is used to recover missing entries using singular value thresholding. Another variation that uses the temporal correlation among channels is shown in \cite{PSCC_algo}.  Since these methods process multiple PMU channels together, they are able to accurately recover missing data sequences on one channel using information from other devices. On the other hand, data prediction accuracy is severely affected by high noise content in any one channel. An online recovery method exploiting the low-rank property of the Hankel matrix constructed by overlapping segments of PMU data has been proposed in \cite{MWang_Hankel}. As columns in a Hankel matrix overlap, this approach faces the following drawbacks- a) repeated entries greatly increase the size of the Hankel matrix, thereby increasing computation burden, b) noise in the matrix elements are highly correlated, c) noisy entries are repeated multiple times affecting recovery accuracy.

To overcome the limitations discussed above, this paper proposes a novel technique for PMU data recovery. A sequence of PMU measurements is first transformed into a \textit{Page matrix} \cite{Page_OG}, and then recovered using a variation of the truncated singular value thresholding algorithm \cite{chatterjeeUSVT2015}. As the matrix columns consist of non-overlapping data segments, they are smaller in size than Hankel matrices and the problem of highly correlated noise in matrix entries is avoided. An optimal hard threshold is used for singular values; thus the matrix rank does not need to be explicitly estimated at every step, resulting in significant computational savings for online prediction. Two variations of the algorithm are proposed, a) an offline imputation method for archived PMU data, and b) an online one-step ahead prediction method. Performance of the Page matrix method is compared to the Hankel matrix method using the same estimation technique. Extensive numerical tests show that both methods have similar recovery accuracy, and the Page matrix method is computationally much faster.

The proposed algorithms can be applied to both univariate and multivariate time-series (both single-channel and multi-channel cases). Although using measurements from multiple PMUs translates to higher accuracy in data recovery, some use cases might warrant processing single channels. For instance, researchers may have access to limited PMU data due to their sensitive nature. Moreover, computation may be sped up by parallelly processing single measurement channels. This may be useful in cases where the streaming synchrophasor data is of superior quality, and significant data drops are not expected.

Contributions of this work may be summarized as follows. \textit{First}, we propose a model-agnostic method for recovering PMU measurements from noisy signals with data drops. The proposed methodology is fast, scalable, easy to implement, and poses minimal memory requirements, making it well-suited for real-time use.  \textit{Second}, the methodology is extended into two algorithms - a) an offline method intended for recovering archived data, and b) an online method for predicting the next measurement, aimed at real-time applications. \textit{Third}, through extensive numerical experiments on simulated and real data, effectiveness of the algorithms is verified. It is shown that the original measurement signals can be reconstructed with high accuracy even in the presence of additive noise and simultaneous data erasures across multiple channels.

The remaining paper is organized as follows. Section \ref{sec:2_a Prelim} describes the missing data recovery problem. Section \ref{sec:2_b algorithms} succinctly presents the online and offline recovery algorithms, and section \ref{sec: 3_experiments} illustrates their performance. Section \ref{sec:4_conclusion} concludes the paper and outlines future research directions.

\section{Problem Set-up}\label{sec:2_a Prelim}
\vspace{-0.1in}
In this section, we establish the mathematical set-up for the PMU data recovery problem and show how it relates to low-rank ME. 
Standard mathematical notations are followed. Calligraphic symbols represent sets, lower case bold letters represent column vectors, and upper case bold letters denote matrices. All zero and all one vectors and matrices of appropriate size are denoted by $\mathbf{0}$ and $\mathbf{1}$ respectively.

\subsection{PMU Data Recovery Using Matrix Estimation}
\vspace{-0.02in}
Simply stated, PMUs are sensors deployed at different points of a power network to measure electrical quantities like voltage and current magnitudes, angles, frequency and rate of change of frequency. Measurements are time-synchronized and typically reported at 30 or 60 frames per second (fps). Due to the physical laws that govern power flow, measurements recorded by a PMU and its neighbors are correlated. Moreover, data within a channel is correlated in time. These spatial and temporal correlations may hence be utilized to recover missing and corrupt measurements. 

Formally, the data recovery problem can be posed as follows. Consider a discrete-time setting with time instants indexed by $t\in\mathds{Z^+}$. Let us say that for each $t\in[1,2,\dots T]$, PMU $i$ records a measurement vector $\boldsymbol{x}_{i}(t)$ of length $c$, where $c$ is the number of measurement channels. Measurements may contain observation noise, and it is assumed that $\mathds{E}[\boldsymbol{x}_i(t)]=\boldsymbol{f}_i(t)$, where $\boldsymbol{f}_i(t)$ are the true values of system states. Although the underlying mean signal $\boldsymbol{f}_i(t)$ is strongly correlated in time, it is assumed that the per-step noise are independent mean-zero random variables with time-varying variance. Given some $\boldx_i(t)$, data recovery algorithms may be designed to address two goals: a) \textit{imputation} (estimate $\boldsymbol{f}_i(t)$ for $t\in[1,2,\dots T]$), and b) \textit{prediction} (estimate $\boldsymbol{f}_i(T+1)$).

Time-series data recovery is a well-studied problem that appears in different domains like econometrics, geosciences and healthcare.
Classical methods for time-series imputation and prediction have employed approaches such as hidden Markov and state-space models \cite{TSA_book}. Different deep neural networks (NN) have also been used \cite{TS_GAN,TS_RNN}. Recent work has shown that low-rank matrix estimation methods can provide simple, effective and computationally efficient means for time-series recovery \cite{TS_matrix,mssa}. This class of methods eliminates the training data requirement of NN models, and hence provide a generalized framework suitable for quick deployment.

The objective of ME is to recover a parameter matrix $\mathbf{M}$ from a partially observed signal matrix $\mathbf{X}$ with corrupt entries, where $\mathds{E}(\mathbf{X})=\mathbf{M}$. A detailed picture of the state-of-the-art is available from \cite{LMRC2,LMRC_review} and references therein. A key observation from ME literature is that matrix $\mathbf{M}$ can be reconstructed from partial and noisy observations by considering a low-rank approximation of the observed matrix. ME algorithms are fairly model-agnostic in terms of the structure of $\mathbf{M}$ and the distribution of $\mathbf{X}$ given $\mathbf{M}$. Therefore, PMU measurements can be transformed into matrices and recovered by applying ME methods to the transformed matrix. Truncated singular value decomposition (SVD) based matrix estimation methods are popularly used \cite{chatterjeeUSVT2015}.

\subsection{Matrix Transformation}\label{subsec:page}
\vspace{-0.02in}

Several methods have been proposed to transform time-series signals into matrices. A naive method involves simply stacking signals together \cite{svd,GEIRI,PSCC_algo}. Although empirically this approach has shown reasonable effectiveness, it cannot be used if very few measurement channels are available. Of course, this is not a pressing concern for a transmission network with many PMUs. But in some cases, it may be necessary to work with a limited number of measurement signals. For instance, researchers outside electric utilities may only have access to limited PMU data. Moreover, it may be desirable to process data from blocks of few PMUs in a parallel manner to speed up computation during real-time application. 

\begin{figure}
    \centering
    \includegraphics[width=0.9\columnwidth]{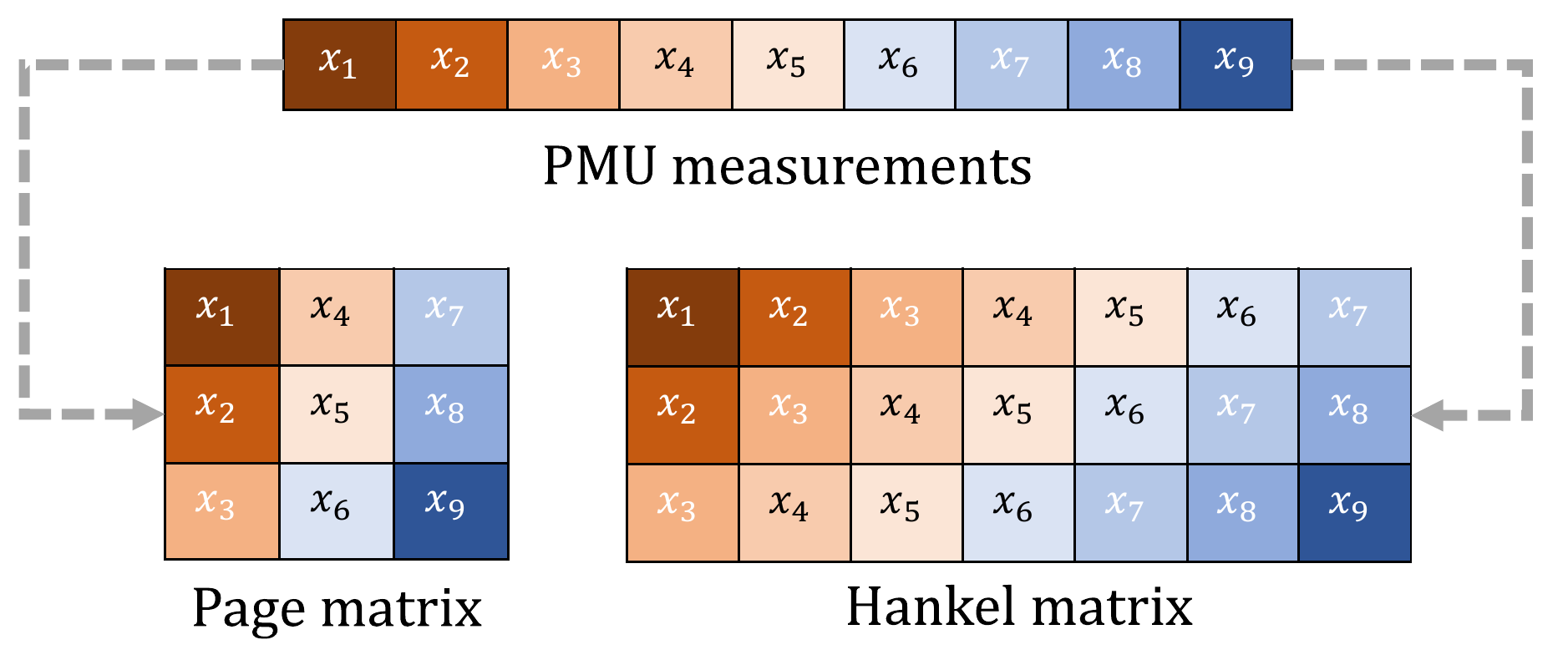}
    \caption{Page and Hankel matrix construction from PMU measurements}
    \label{fig:matrixify}
    \vspace{-0.25in}
\end{figure}

An alternative approach described in \cite{MWang_Hankel} uses the Hankel matrix transformation. Overlapping segments of PMU data are placed side by side to form a Hankel matrix. That such matrices are approximately low-rank has been empirically verified in \cite{MWang_Hankel}. As the Hankel matrices contain repeated entries, they are large in size and noisy elements appear multiple times. Repetition of noisy entries may reduce data recovery accuracy and the large matrix size increases computation burden. 

The limitations of Hankel matrices may be overcome by using Page matrices \cite{TS_matrix, mssa}. A Page matrix is constructed from observation vector $\boldx(t)$ by placing contiguous segments of size $L>1$ side by side as non-overlapping columns of the resultant matrix \cite{Page_OG}. A schematic description of the matrix transformation process is shown in fig. \ref{fig:matrixify}.
It can be seen that when a time-series of the same length is transformed, the Hankel matrix is much larger than the corresponding Page matrix. The low-rank property of Page matrices has also been examined in \cite{TS_matrix}. It is established that, in expectation, for a large class of processes, the generated Page matrix is either exactly or approximately low-rank. These proceeses include linear recurrent functions (LRF) described by $f(t)=\sum_{g=1}^G \alpha_g f(t-g)$, for some $G\geq1$. That power systems quantities such as bus voltages, line currents and frequency follow LRF has been previously concluded in literature. For example, it is posited that steady state power systems measurements at any instant are a linear combination of the last three measurements \cite{PMU_AR}. This result has also been used to deisgn the kalman filter-based data prediction and smoothing algorithm in \cite{kalman}.

The LRF nature of PMU measurements also helps in formulating a prediction algorithm. Once the low-rank approximation of the transformed Page matrix is obtained via some ME technique, the last row of this matrix can be expressed as a linear combination of the other rows. Therefore, future values can be forecast by applying linear regression to the approximate Page matrix.

\subsection{Optimal Singular Value Thresholding (OSVT)} \label{subsec:OSVT}

Various techniques exist for low-rank ME \cite{LMRC_review,LMRC2}. In the proposed model-agnostic recovery framework, no information about the rank of the Page matrices is available beforehand. Hence, a variation of the truncated SVD method that does not need matrix rank information has been used in this paper \cite{chatterjeeUSVT2015}. The optimal singular value threshold is chosen based on findings reported in \cite{optimal_singular_value}. The main steps in the estimation algorithm are detailed next in algorithm \ref{algo:OSVT}. It is assumed that missing data points in the observation matrix are preliminarily filled in by the last available observation. 

\begin{algorithm}
\caption{Optimal Singular Value Thresholding (OSVT)}\label{algo:OSVT}
\begin{algorithmic}[1]
\State \textbf{Scaling: } Entries of the observation matrix $\mathbf{X}$ are scaled to lie in the interval -1 to 1. Let the scaled observation matrix be called ${\mathbf{Y}}$ with individual entries $y_{i,j}$. Mathematically, ${y}_{i,j} = ({x}_{i,j}-0.5(a+b))/0.5(b-a)$, where $a$ and $b$ are the minimum and maximum entries of $\mathbf{X}$ respectively.
\State \textbf{Singular value decomposition: } Let ${\mathbf{Y}}=\sum_{i=1}^m \sigma_i\boldsymbol{u}_i\boldsymbol{v}_i^T$ be the singular value decomposition of ${\mathbf{Y}}$. The singular values are given by $\sigma_i$; and $\boldsymbol{u}_i$ and $\boldsymbol{v}_i$ are the corresponding left and right singular vectors respectively. 
\State \textbf{Singular value thresholding: } Choose a set $S$ of \textit{thresholded} singular values such that : $S:= \{\sigma_i>\sigma_{th}\}$, where the optimality threshold $\sigma_{th}$ is given by:
\[\sigma_{th} = \sqrt{2(\zeta+1)+\frac{8\zeta}{(\zeta+1)+\sqrt{\zeta^2+14\zeta+1}}}\]
Here, $\zeta = m/n$, where ${\mathbf{Y}}$ is a $m\times n$ matrix. Moreover, $m\leq n$. In the case that $m>n$, the estimation algorithm must be applied to $\mathbf{X}^T$ to obtain an estimate of $\mathbf{M}^T$. 
\State \textbf{Low rank approximation: } The low rank approximation of matrix $\mathbf{Y}$ is given by $\mathbf{\hat{Y}}=\sum_{\sigma_i\in S} \sigma_i \boldsymbol{u}_i \boldsymbol{v}_i^T$. The final estimation $\hat{\mathbf{M}}$ of the parameter matrix is obtained by scaling back the values of $\hat{\mathbf{Y}}$ to the interval $[a,b]$.
\end{algorithmic}
\end{algorithm}

The hard singular value threshold proposed in \cite{optimal_singular_value} is optimal in an asymptotic sense. It is postulated that for large low-rank matrices, when a data singular value $\sigma_i$ is too small, the corresponding singular vectors $\boldsymbol{u}_i$ and $\boldsymbol{v}_i$ are very noisy and the component $\sigma_i \boldsymbol{u}_i \boldsymbol{v}_i^T$ should not be used in approximating matrix $\hat{\mathbf{Y}}$ from $\mathbf{Y}$. The cutoff for singular values is determined to be $\sigma_{th}$ as described in algorithm 1. The alternative method used in literature for estimating matrix rank involves- a) selecting a threshold for rank approximation error, and b) choosing the lowest rank for which the approximate matrix does not violate the predetermined error threshold. The choice of the approximation error threshold is somewhat arbitrary and has been empirically decided in works like \cite{MWang_Hankel}. Using a hard threshold eliminates the need for repeated calculations of rank approximation error at every step, thereby significantly improving computation speed. It is further shown in \cite{optimal_singular_value} that the optimally tuned thresholding method outperforms (in terms of mean squared error) classical truncated SVD when signal noise content is low to moderate; and the methods perform roughly similarly when noise content is high.

\subsection{Multivariate Time-Series Recovery} \label{subsec:stacked_page}
\vspace{-0.02in}
\begin{figure}
\vspace{-0.06in}
    \centering
    \includegraphics[width=0.89\columnwidth]{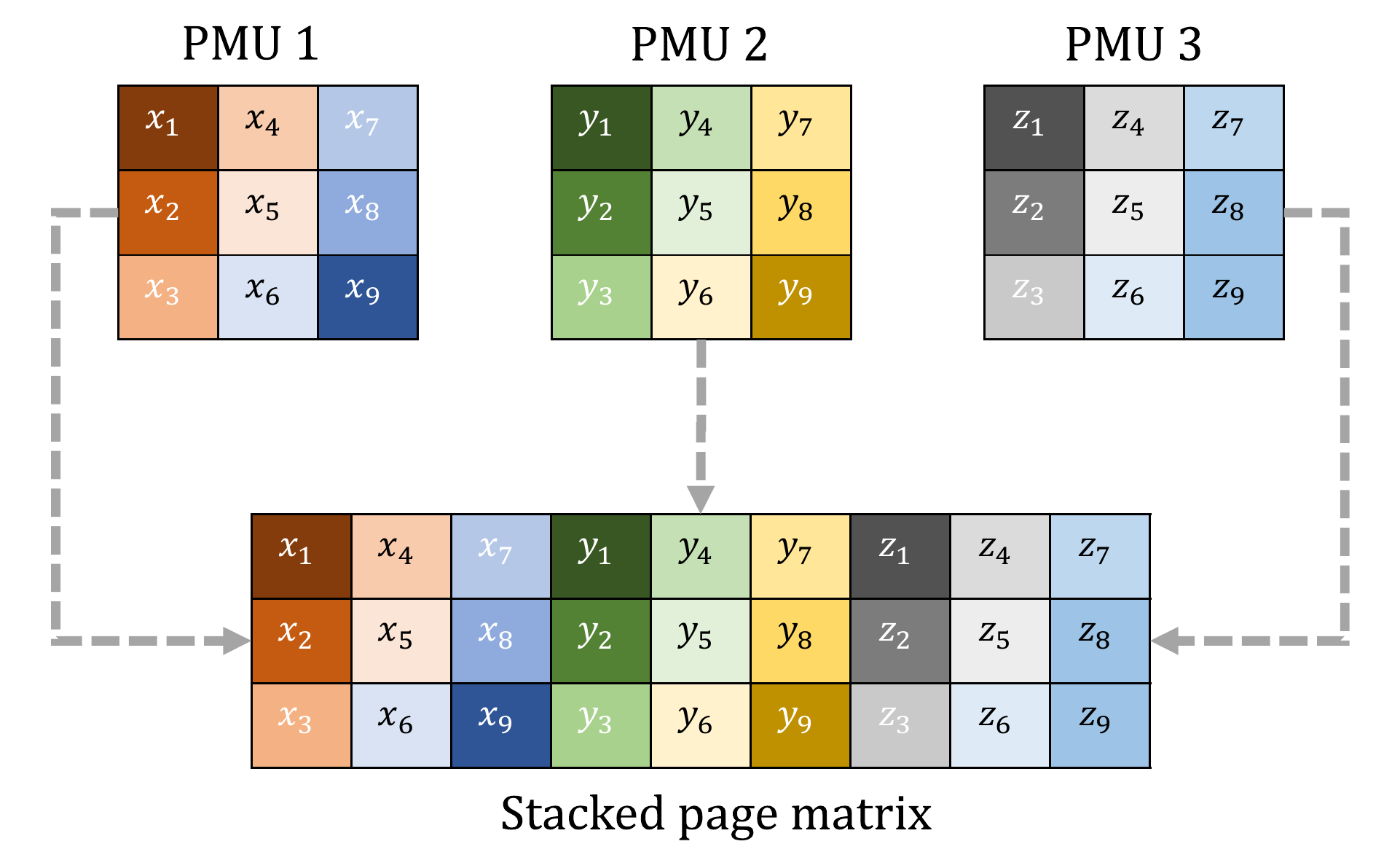}
    \caption{Stacked Page matrix for multivariate data recovery}
    \label{fig:stacked_matrix}
    \vspace{-0.25in}
\end{figure}

As mentioned before, PMU measurements in a network are correlated, and hence information from different PMUs can be utilized to recover degraded data. Readings from multiple PMU channels can be transformed into a `stacked' Page matrix by concatenating individual Page matrices columnwise, as shown in fig. \ref{fig:stacked_matrix}. The low-rank property of stacked Page matrices for a large class of processes including LRF has been verified in \cite{mssa}.
Let $N$ be the total number of PMU measurement channels available. The data recovery process can now be summarized with the following steps: a) transform observation vectors $\boldsymbol{x}_i(t), i\in[1,2,\dots, N], t\in[1,2,\dots,T]$ to Page matrices $\mathbf{P}_i, i\in[1,2,\dots,N]$, each with dimension $L\times T/L$, by placing non-overlapping contiguous data segments of length $L>1$ (an algorithmic hyperparameter) side-by-side as columns; b) concatenate the individual Page matrices columnwise to form a stacked Page matrix $\mathbf{X}$ of dimension $L\times (NT/L)$; c) perform OSVT (algorithm 1) on matrix $\mathbf{X}$ to obtain a denoised estimate $\hat{\mathbf{X}}$; d) learn a forecasting model that expresses the last row of $\hat{\mathbf{X}}$ as a linear combination of its remaining rows. The in-sample imputation and prediction error for this method scales as $1/\sqrt{NT}$ \cite{mssa}.

\section{Data Recovery Algorithms} \label{sec:2_b algorithms}
\vspace{-0.06in}
This section shows how the individual pieces described in the previous section are put together for recovering degraded PMU data. The algorithms can be applied to both single-channel and multi-channel data using suitable Page matrices.

\subsection{Offline Data Imputation}
\vspace{-0.02in}
Let us consider the problem of denoising already recorded PMU measurements and imputing missing readings. Assume $n$ number of PMUs, each with $c$ measurement channels. Then the total number of data channels available is $N=n\times c$.
{The observations are partitioned into say $k$ windows of length $T$ each. Note here that for the imputation task, hours of data can be processed at once. Thus, $T$ can be quite large. It is assumed here that $T$ is perfectly divisible by the value of $L$ chosen.}
The sequential steps to be performed for offline data imputation are listed in algorithm 2. 

\begin{algorithm}[h!]
\caption{Offline PMU data imputation}\label{algo:offline}
\begin{algorithmic}[1]
\State{\textbf{Initialization:} Set $T\leftarrow$ window length, $N\leftarrow$ number of PMU channels, $k\leftarrow$ number of data windows.}
\State {\textbf{For} $j = 1:k$, \textbf{do}}
\State \hspace{8pt} {\textbf{Matrix transformation:} Construct a stacked Page matrix using data from $N$ PMU channels in the $j^{th}$ measurement window with $T$ observations, as described in section \ref{subsec:stacked_page}.} 
\State\hspace{8pt} {\textbf{Low-rank matrix estimation:} Obtain a low-rank approximation of the stacked Page matrix $\hat{\mathbf{X}}$ using the OSVT method outlined in algorithm \ref{algo:OSVT}.} 
\State\hspace{8pt}{\textbf{Recover estimated measurements:} Reshape matrix $\hat{\mathbf{X}}$ to recover the estimated measurements.}
\State {\textbf{End for}}
\end{algorithmic}
\end{algorithm}

\subsection{Online Data Prediction}
\vspace{-0.02in}
The online forecast problem pertains to predicting the signal value $\boldsymbol{f}(T+1)$ given past observations $\boldsymbol{x}(t), t\in[1,2,\dots,T]$. This is analaogous to performing regression with noisy data. The online forecasting algorithm proposed in this paper- \textit{first}, denoises and imputes past observations (algorithm \ref{algo:offline}), and \textit{second}, uses linear regression to learn the relationship between the last row and remaining rows of the imputed observation matrix. Next, the learned regression parameters are used to predict the next sample from a Page matrix shifted by one sample. Of course, the first data window of length $T$ needs to be filled before next-step prediction can proceed. The forecast procedure is described with better clarity in algorithm \ref{algo:online}. 

\begin{figure} [t!]
    \centering
    \includegraphics[width=0.8\columnwidth]{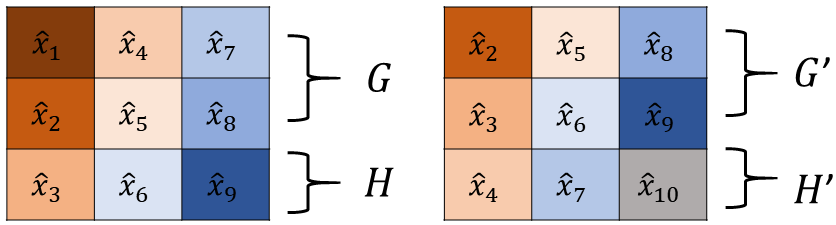}
    \caption{Visual description of matrices $\mathbf{G},\mathbf{H}, \mathbf{G'}$ and $\mathbf{H'}$ for a Page matrix using data from a single channel}
    \label{fig:reg}
     \vspace{-0.1in}
\end{figure}

\begin{algorithm}[h!]
\caption{Online PMU data prediction}\label{algo:online}
\begin{algorithmic}[1]
\State {\textbf{Initialization:} Set $T \leftarrow$ window length, $N \leftarrow$ total number of PMU channels, $j\leftarrow 0$}.
\State {\textbf{While} PMU data streams are available, \textbf{do}:}
\State \hspace{8pt} {\textbf{Matrix formation:} Convert measurement vectors $\boldsymbol{x_i}(t+j)$, $i\in[1,2,\dots N]$, $t\in[1,2,\dots T]$ to a stacked Page matrix $\mathbf{X_{T+j}}$, say. } 

\State \hspace{8pt} \textcolor{black}{\textbf{Matrix imputation:} 
Using algorithm \ref{algo:offline}, denoise and impute $\mathbf{X_{T+j}}$ to obtain ${\mathbf{\hat{X}_{T+j}}}$. }

\State \hspace{8pt} {\textbf{Learn the linear forecast model:} Partition matrix $\mathbf{\hat{X}_{T+j}}$ into two parts $\mathbf{G}$ and $\mathbf{H}$ such that $\mathbf{G}$ comprises of the first $L-1$ rows in $\mathbf{\hat{X}_{T+j}}$ and $\mathbf{H}$ contains the last. Linear regression here pertains to estimating the parameter vector {$\boldsymbol{\beta_j}$} for which $\mathbf{H}=\mathbf{G}{\boldsymbol{\beta_j}}+\boldsymbol{\epsilon}$ in the least squares sense. }

\State \hspace{8pt} {\textbf{Forecast the one-step-ahead data:} Construct matrix $\mathbf{G'}$ with the last $L-1$ rows of $\mathbf{\hat{X}_{T+j}}$. Estimate $\mathbf{H'}$ from $\mathbf{G'}$ as $\mathbf{H'}=\mathbf{G'}\boldsymbol{\beta_j}$ using the value of $\boldsymbol{\beta_j}$ learnt in step 5.
For the univariate prediction case, the last entry of $\mathbf{H'}$ is the prediction of the next measurement $x(T+j+1)$. For the multivariate case, pertinent entries from $\mathbf{H'}$ need to be extracted. For clarity, a visual description of the matrices $\mathbf{G},\mathbf{H}, \mathbf{G'}$ and $\mathbf{H'}$ is provided in fig. \ref{fig:reg}}.

\State \hspace{8pt} {\textbf{Update:} Set $j\leftarrow j+1$}
\State {\textbf{End while}}
\vspace{-2pt}
\end{algorithmic}
\end{algorithm}

\begin{figure*}[h!]
\centering
    \begin{minipage}[t]{0.32\textwidth}
    \centering
    \vspace{0pt}
    \includegraphics[width=0.9\columnwidth]{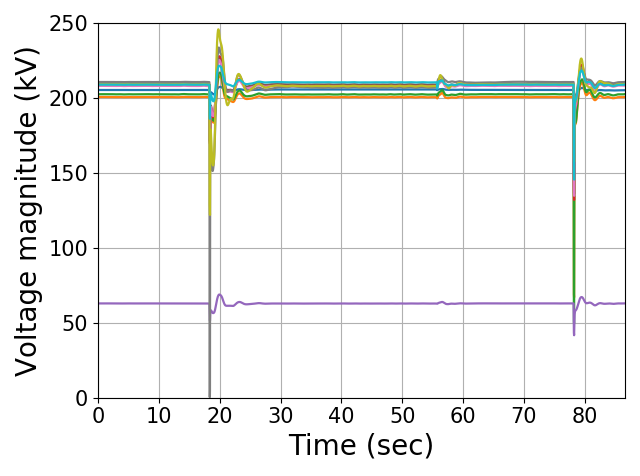}
    \caption{PMU measurements: Positive sequence voltage magnitude}
    \label{fig:PMU_V}
    \end{minipage}%
    \begin{minipage}[t]{0.32\textwidth}
    \centering
    \vspace{0pt}
    \includegraphics[width=0.9\columnwidth]{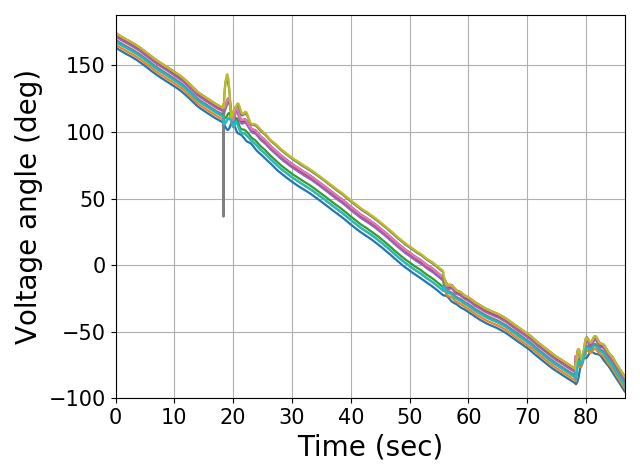}
    \caption{PMU measurements: Positive sequence voltage angle (unwrapped)}
    \label{fig:PMU_A}
    \end{minipage}%
    \begin{minipage}[t]{0.32\textwidth}
    \centering
    \vspace{0pt}
     \includegraphics[width=0.9\columnwidth]{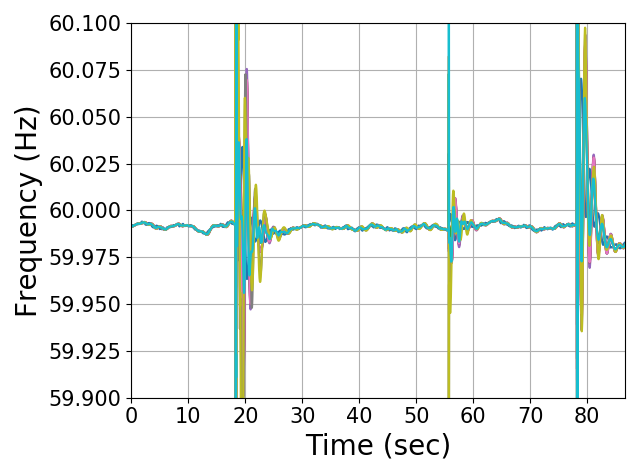}
    \caption{PMU measurements: Frequency}
    \label{fig:PMU_F}
    \end{minipage}%
    \vspace{-0.2in}
\end{figure*}

\subsection{Choice of Hyperparameters}

{Selecting \textit{good} hyperparameters is essential for achieving high-accuracy data recovery while minimizing computation time. For the data recovery algorithms proposed in this work, two main hyperparameters need to be chosen- a) $L$ or the number of rows in the Page matrix, and b) $T$ or length of data window. }

{$\bullet$\textit{ Choice of $L$}: Empirically, it was observed that choosing a $L$ value in the 5-10 range yielded good results for both the online prediction and offline imputation tasks. Keeping the parameter $L$ small enables capturing the short-term temporal patterns in PMU data. However, for very noisy data, increasing $L$ can help in obtaining smoother estimates.}

{$\bullet$ \textit{Choice of $T$}: The in-sample imputation and prediction error for the multivariate data recovery method scales as $1/\sqrt{NT}$. 
Therefore, the longer the data window, the better the prediction accuracy. On the other hand, choosing a long data window increases computation time. As computation time is not a prime concern for offline data imputation, a fairly long window can be chosen. In the numerical results section of this paper, data imputation with $T=54000$ (30 minutes of data) has been demonstrated .}


{Computation time is of greater concern when it comes to online prediction, as every prediction step requires a matrix estimation and linear regression operation. Hence, choosing a shorter time window is beneficial. However, the window length $T$ should be selected carefully. If $T$ is too small, prediction accuracy will be impacted by measurement noise. If $T$ is too large, the data window may contain obsolete modes thereby degrading prediction accuracy. It was empirically observed that a data window of about $\sim$ 30-45 samples yielded good performance without unduly increasing prediction time.}

{\subsection{Scalability} The size of the Page matrix will also be determined by the number of PMU channels available. In a real system with hundreds of field PMUs, computation may be sped up by dividing the PMUs into groups with similar modal signatures and processing the groups in parallel. Grouping together PMUs geographically close to each other will also enhance visibility into `local' dynamics which might have been obscured by aggregating signals over a wide-area grid. Some strategies for grouping PMU signals to ensure low-rankness of the measurement window have been presented in literature \cite{PMU_grouping}. }

During a disturbance, the system deviates from its predicted behavior, and the low-rank property of the stacked Page matrix may not hold true. Therefore, at the onset of a disturbance, online predictions may vary greatly from observed measurements. The difference between the actual observations and algorithm predictions decreases gradually.

\section{Numerical Results} \label{sec: 3_experiments}
\vspace{-0.06in}
This section describes the numerical tests performed to validate the performance of the proposed imputation and prediction methodologies. The first set of tests are performed on simulated data, artificially distorted by the injection of additive noise and random data drop. The next set of tests considers real noisy PMU measurements recorded by a U.S. utility. in this case, the \textit{true values} of the measurements are unknown, but visual inspection reveals that reasonable values are predicted for swathes of missing data. All computations are performed on a PC with 16 GB RAM and 2.6 GHz Intel core i7-9750HF processor.

\subsection{Simulated Measurements Dataset}
\vspace{-0.04in}
Numerical tests were performed on 86.6 seconds of measurements from PMUs installed at ten generator bus terminals of the IEEE 39-bus transmission model\cite{dataport}.  Quasi-steady state operations and three-phase faults were recorded using RTDS power systems simulator and GTNETx2 based PMUs at 60 fps reporting rate. During the length of the simulation, quasi steady-state conditions were simulated by modulating the mechanical torque of generator $G_1$ every 200 ms by a random perturbation within $\pm 1\%$ of the nominal value. The data also shows three disturbances. At 18.33 seconds, a self-clearing three-phase fault is followed by tripping of the faulted line, leading to a topology change. At 55.67 seconds, the tripped line is reconnected, restoring the initial network topology. At 78.13 seconds, another three-phase self-clearing three-phase fault takes place.

In this study, three measurement channels from each PMU were used - positive sequence voltage magnitude, positive sequence voltage angle and frequency. Data from all ten PMUs are shown in figures \ref{fig:PMU_V}, \ref{fig:PMU_A} and \ref{fig:PMU_F}. PMUs are referred to by the generator terminal they are installed at. For example, the PMU at generator $G_1$ terminal is called PMU $G_1$. Voltage angle at PMU $G_1$ is considered as the reference angle. 

Simulated measurements have been used for evaluating the data recovery algorithms as the `\textit{ground truth}' data is available for comparison. On the other hand, in real PMU measurements, some readings may already be missing or corrupt, and there is no way of \textit{exactly knowing} what those measurements should have been. Further, as this simulation records network topology changes, we can investigate if varying topologies affect recovery accuracy.

\subsection{Data transformation}
\vspace{-0.05in}
The PMU measurements are scaled before being transformed into the stacked Page matrix described in section \ref{subsec:stacked_page}. The scaling process used in this paper is described below:\\
$\bullet$ \textit{Voltage magnitude}: Measurements were transformed into the per unit (p.u.) system.\\
$\bullet$ \textit{Voltage angle}: The reference voltage angle is subtracted from individual channel data. In the dataset used, angle readings were already unwrapped. For unwrapping angles in real-time, the strategy outlined in \cite{unwrap_angle} may be followed.\\
$\bullet$ \textit{Frequency}: Frequency measurements were scaled as follows: $f_{scaled} = (f_{measured}-60)\times 10 $.

The scaling method described above empirically showed good results; however other approaches may also be used. 

\subsection{Error Metric}\vspace{-0.04in}
Mean absolute percentage error (MAPE) has been used as the error metric to evaluate the accuracy of the proposed algorithms. Mathematically, MAPE for a time-series of length $n$ maybe expressed as:
\begin{align*}
    MAPE = \frac{1}{n}\sum_{t=1}^n \bigg| \frac{A_t-\hat{A_t}}{A_t} \bigg|
\end{align*}
Here, $A_t$ is the actual time-series value and $\hat{A_t}$ is the corresponding prediction. As it intuitively conveys relative error, MAPE is widely used in regression problems and model evaluation tasks.

\begin{figure}
    \centering
    \includegraphics[width=0.8\columnwidth]{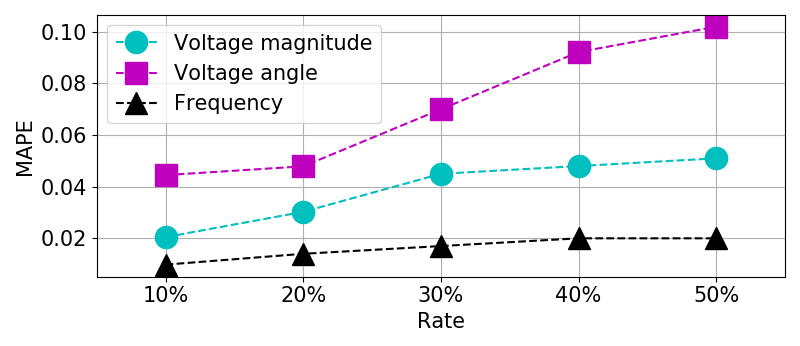}
    \caption{Error in $G_2$ PMU channels with simultaneous data drop in all PMU channels (median over 20 runs)}
    \label{fig:MAPE_drop}
    \includegraphics[width=0.8\columnwidth]{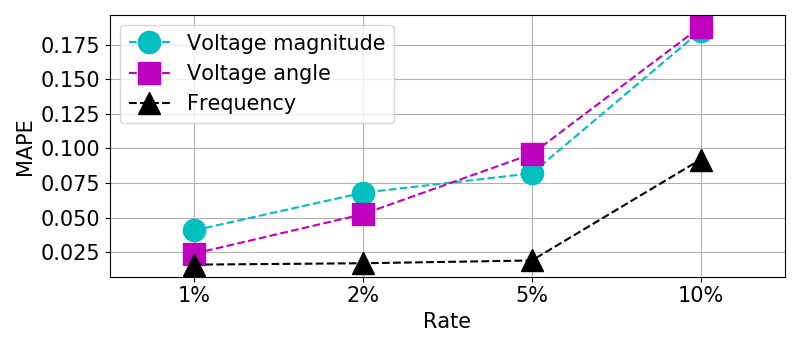}
    \caption{Error in $G_2$ PMU channels with zero-mean additive noise on all $G_2$ PMU channels (median over 20 runs)}
    \label{fig:MAPE_noise}
    \vspace{-0.2in}
\end{figure}

\subsection{Offline Data Imputation}\label{subsec:offline_data_impu}

\begin{figure*}
    \centering
    \begin{minipage}[t]{0.33\textwidth}
    \subfloat[Positive sequence voltage magnitude]{\includegraphics[width=\textwidth]{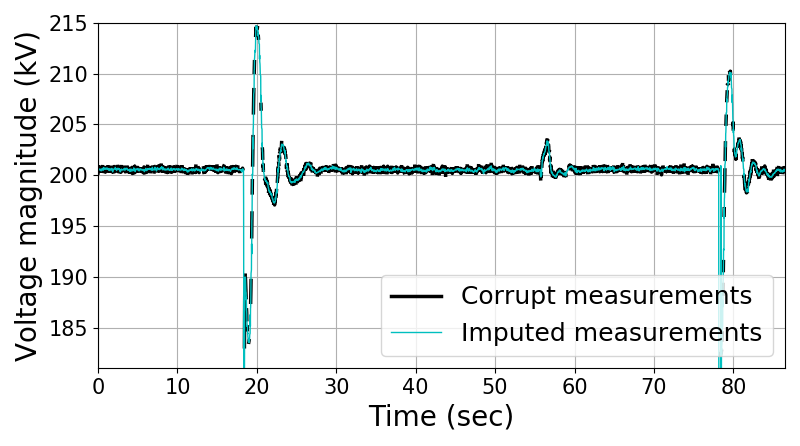}}\\\vspace{-0.1in}
    \end{minipage}%
    \begin{minipage}[t]{0.33\textwidth}
    \centering
    \subfloat[Positive Sequence voltage angle (referenced)]{\includegraphics[width=\columnwidth]{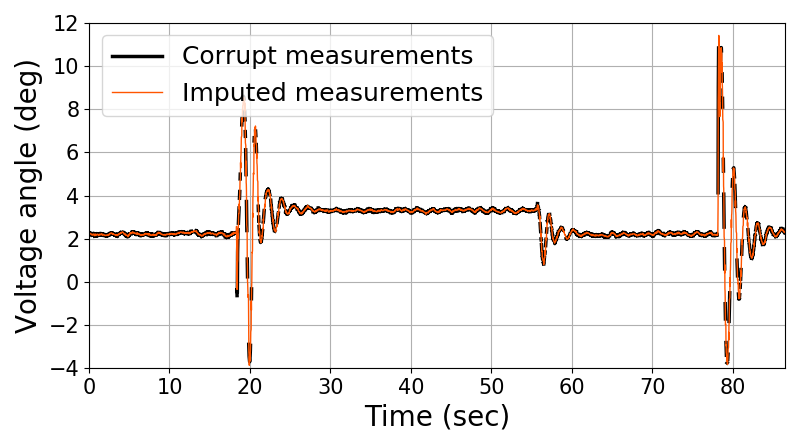}}\\\vspace{-0.1in}
    \end{minipage}%
    \begin{minipage}[t]{0.33\textwidth}
    \centering
    \subfloat[Frequency]{\includegraphics[width=\columnwidth]{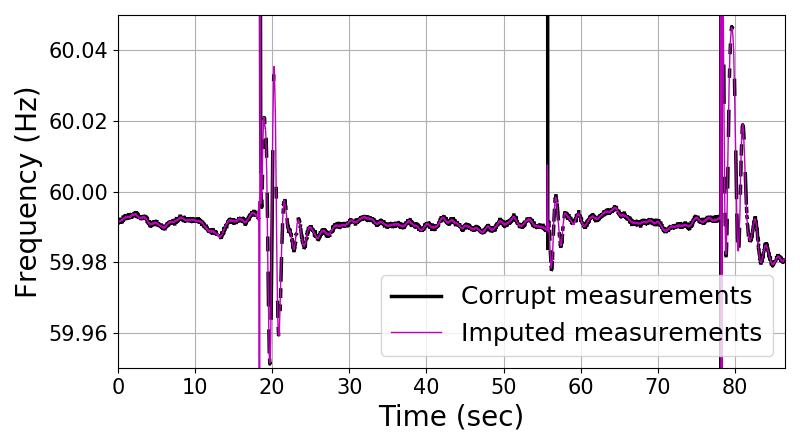}}
    \end{minipage}%
    \vspace{0.04in}
    \caption{Imputed measurements with 50\% simultaneous missing data and 2\% additive noise on all $G_2$ PMU channels }
    \label{fig:imputed_G2}
    \begin{minipage}[t]{0.33\textwidth}
    \centering
    \subfloat[Positive sequence voltage magnitude]{\includegraphics[width=\columnwidth]{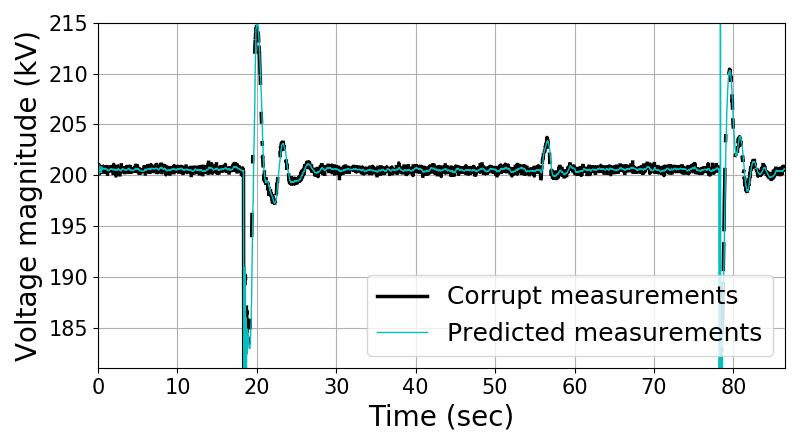}}\\
    \end{minipage}%
    \begin{minipage}[t]{0.33\textwidth}
    \centering
    \subfloat[Positive Sequence voltage angle (referenced)]{\includegraphics[width=\columnwidth]{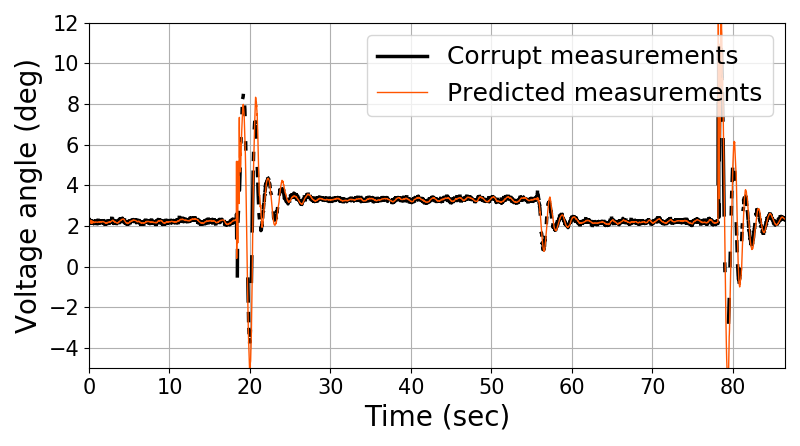}}\\
    \end{minipage}%
    \begin{minipage}[t]{0.33\textwidth}
    \subfloat[Frequency]{\includegraphics[width=\columnwidth]{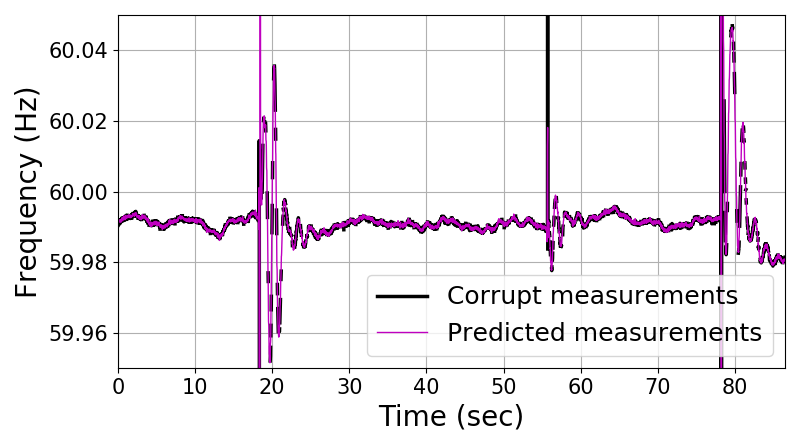}}
    \end{minipage}
    \vspace{0.04in}
    \caption{Predicted measurements with 50\% simultaneous missing data and 2\% additive noise on all $G_2$ PMU channels }
    \label{fig:predicted_G2}
    \vspace{-0.2in}
\end{figure*}

As communication channel congestion may impact all PMU channels, data erasures may be correlated. To capture the efficacy of the proposed imputation method under realistic data degradation conditions, the following modes were checked:

$\bullet$ \textit{Data drop}: Simultaneous data drops on all PMU channels in the network were considered. A fraction of timestamps (determined by the chosen data drop rate) were randomly selected and corresponding measurements were dropped from all PMU channels. As the data drop rate is increased, the chance of missing consecutive data segments also increases. Error on all channels of PMU $G_2$ as the data drop rate is varied is shown in fig. \ref{fig:MAPE_drop}. These are median values of observations over 20 runs. It is evident that the measurements can be reconstructed with acceptable accuracy. Time taken for imputation did not vary significantly with data drop rate and median time taken over 100 runs was 0.0184 seconds. In the interest of brevity, only results for PMU $G_2$ has been included in this paper, but similar results were obtained for other PMU channels as well. 

Results for the extreme scenario where readings from all PMUs are missing are shown here. For \textit{less extreme} cases, i.e. when data from only some of the PMU channels are missing, higher accuracy in signal reconstruction may be expected.

$\bullet$ \textit{Additive noise}: 
{Noise in PMU measurements may arise due to errors in calibration, instrumentation and quantization. Existing studies suggest that the zero-mean gaussian distribution is a suitable model to characterize this noise. The signal-to-noise ratio (SNR) for real transmission-level PMUs is estimated to be around $\sim$45 dB, while for distribution-level PMUs the SNR is estimated to be lower \cite{PMU_noise}. Similar noise was injected into PMU $G_2$ channels for the numeric tests in this work. }

Zero-mean gaussian noise was added to all PMU $G_2$ channels. The standard deviation of the noise distribution on each channel was given by a percentage of the median of true steady-state data (let us call this percentage the \textit{noise rate}). Data recovery error with varying noise rates is shown in fig. \ref{fig:MAPE_noise}. Median computation time over 100 runs was 0.0192 seconds. 

For both the cases above, the number of rows in the stacked Page matrix was 10. Fig. \ref{fig:imputed_G2} shows both the corrupt and imputed data when 2\% noise was added to the PMU $G_2$ channels and 50\% of the readings were missing. It can be seen that the PMU signals are reconstructed with reasonable accuracy even when consecutive data segments are missing. Further, the data recovery accuracy is robust to topology changes in the power network.

\begin{figure*}
    \centering
    \includegraphics[width=0.8\textwidth]{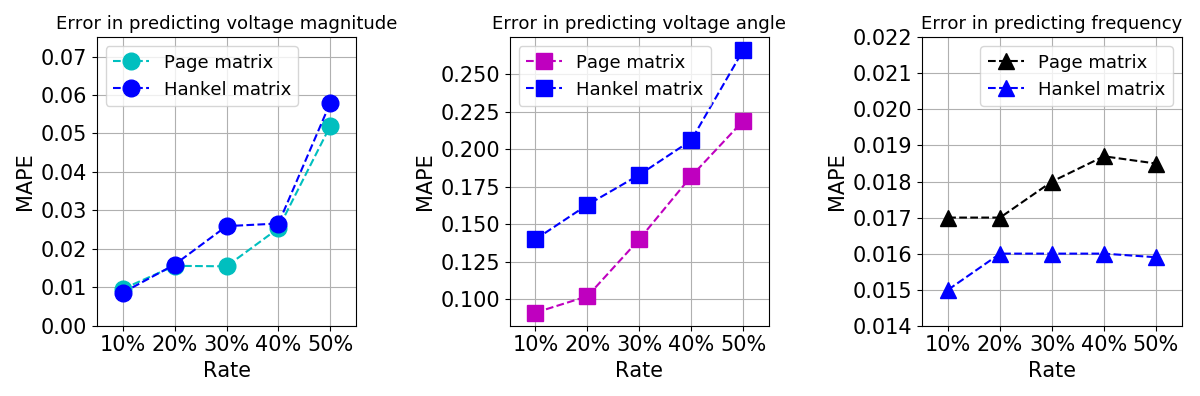}
    \caption{Prediction error in $G_2$ PMU channels with simultaneous missing data on all $G_2$ PMU channels (median over 20 runs)}
    \label{fig:pred_missing}
    \includegraphics[width=0.8\textwidth]{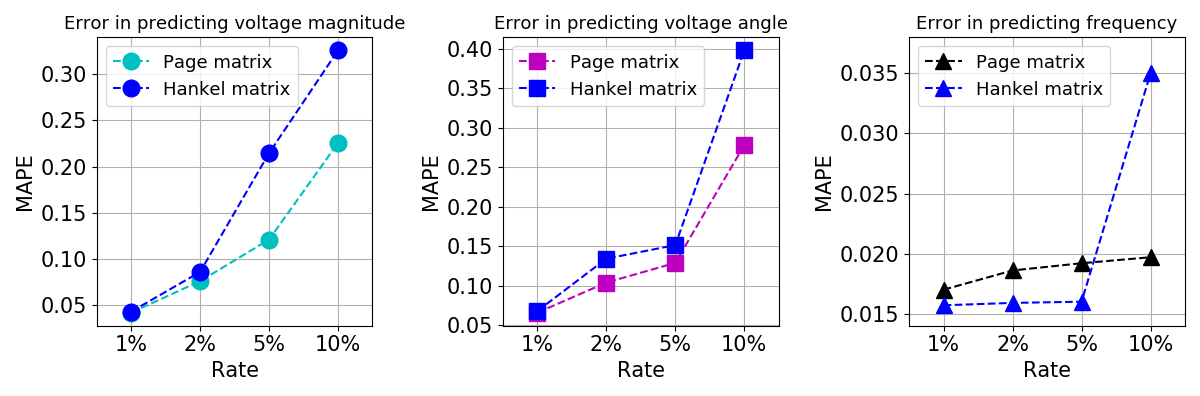}
    \caption{Prediction error in $G_2$ PMU channels with additive noise on all $G_2$ PMU channels (median over 20 runs)}
    \label{fig:pred_noise}
    \vspace{-0.1in}
\end{figure*}

\subsection{Online Data Prediction}
Similar numerical tests were conducted with the same PMU dataset for assessing online data prediction accuracy. Performance of the stacked Page-matrix based prediction algorithm has been compared with the Hankel-matrix based prediction method put forth in \cite{MWang_Hankel}. For better comparison, the same ME technique is used for both algorithms.

$\bullet$ \textit{Prediction error}: Figures \ref{fig:pred_missing} and \ref{fig:pred_noise} show the prediction error for different data drop rates (simultaneous) and additive noise on all $G_2$ PMU channels. The results are median observations over 20 runs. Data drops and noise were introduced in the data in the same manner as discussed in section \ref{subsec:offline_data_impu}. It is observed that the prediction error of the Page matrix method is similar to/slightly better than the Hankel matrix based method. The corrupt and reconstructed signals when 50\% data drop and 2\% noise is added to all $G_2$ PMU channels is shown in fig. \ref{fig:predicted_G2}. In the experiments, number of rows used was 5, and window length considered was 25.

$\bullet$ \textit{Prediction time}: One-step ahead predictions have multiple applications. Missing samples can be filled in with predicted values. Similarly, irregularities in data may be detected looking at how far  measurements stray from their predictions. Now, for any real-time algorithm implementation, computation time is of prime concern. U.S. electric utilities typically use PMUs with reporting rates of 30 or 60 fps. For these PMUs, the time intervals between two consecutive samples is 0.0333 or 0.0167 seconds. Therefore, in order to predict a sample before it arrives at the control center, the prediction time must be much lower.  

Our numeric tests showed that the prediction time for the proposed Page matrix algorithm was $\sim$0.001 seconds. In comparison, time taken by the Hankel matrix based method was $\sim$0.003 seconds. All times are median values recorded over 100 runs. The prediction times did not vary significantly with varying rates of data drop or noise. Thus, we see that the method proposed in this paper is much faster than the Hankel-matrix based method, and provides similar/slightly better accuracy.
 
{$\bullet$ \textit{Window length of Hankel matrix:} The preceding set of experiments showed that the proposed Page matrix-based data recovery method provided accuracy levels similar to the Hankel matrix-based method, while speeding up computations. The computational savings are largely due to the smaller size of the Page matrix for the same measurement window. A natural question arises here: how would using a shorter time window for the Hankel matrix method affect data recovery accuracy? It is expected that predictions obtained using a smaller number of observations will be noisy and less accurate. This notion is experimentally verified in this work.}
\begin{figure*}
\centering
\includegraphics[width=\textwidth]{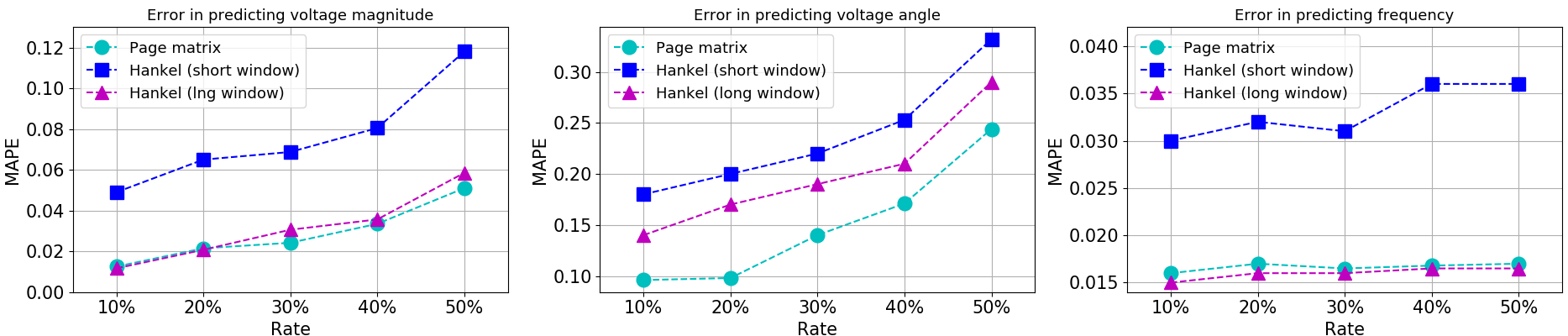}
\caption{\cmb{Prediction error in $G_2$ PMU channels with simultaneous missing data on all $G_2$ PMU channels (median over 20 runs)}}
\label{fig:Hankel_long_short}
\vspace{-0.15in}
\end{figure*}

{Figure \ref{fig:Hankel_long_short} shows how the recovery accuracy varies when data is dropped simultaneously from all channels of PMU $G_2$ using the same methodology as section \ref{subsec:offline_data_impu}. Three matrix transformations were checked, a) Page matrix (dimensions $5\times 6$), b) Hankel matrix with longer window (dimensions $5\times 26$), and c) Hankel matrix with shorter window (dimensions $5\times 6$). For the first two cases, measurement window length used for predicting the next sample is $T=30$, while for the third case it becomes $T=10$. As the Page matrix and Hankel matrix with shorter window have the same dimensions, time taken to predict the next sample using these matrices is almost the same. However, as evident from figure \ref{fig:Hankel_long_short}, the Page matrix method yields higher recovery accuracy. When using a Hankel matrix with longer window, the prediction accuracy improves, but computation time increases as well.}

\begin{figure}
    \centering
    \includegraphics[width=\columnwidth]{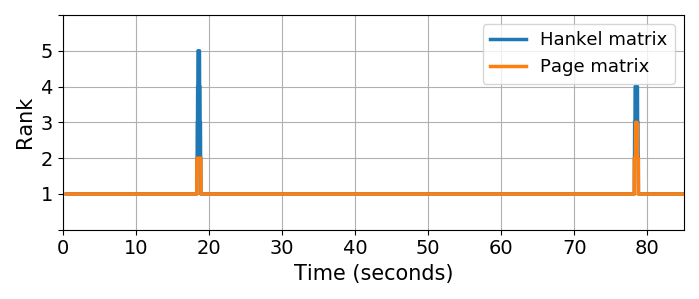}
    \caption{Matrix rank for different data windows}
    \label{fig:ranks}
    \vspace{-0.2in}
\end{figure}

$\bullet$ \textit{Verifying the low rank-property}: The low rank properties of both the stacked Page and Hankel matrices were checked for all the data windows, as shown in fig. \ref{fig:ranks}. It can be seen that both the Page and Hankel matrices are generally low-rank, and the matrix ranks increase at the beginning of events. During the first event, the Hankel matrix becomes full-rank, and the low-rank property does not hold.

\subsection{Real PMU measurements dataset} \vspace{-0.05in}
Next, we check how well the proposed algorithms perform with real PMU data from an anonymized U.S. electric utility. The data corresponds to 30 minutes of measurements (54000 samples) from four PMUs of the utility, each of which reports three channels- positive sequence voltage magnitude, positive sequence voltage angle, and frequency. PMU reporting rate is 30 fps. The voltage angle of PMU4 is considered as reference, since it has the least amount of missing entries. Voltage angles have been unwrapped using the algorithm from \cite{unwrap_angle}.

\begin{figure}
    \centering
    \includegraphics[width=\columnwidth]{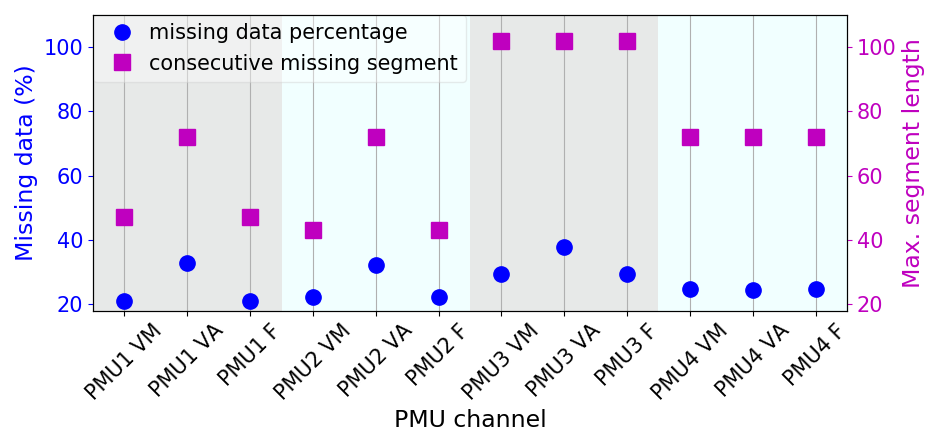}
    \caption{Data quality in the real PMU dataset}
    \label{fig:dom data quality}
    \vspace{-0.2in}
\end{figure}

\begin{figure*}
\centering
\includegraphics[width=\textwidth]{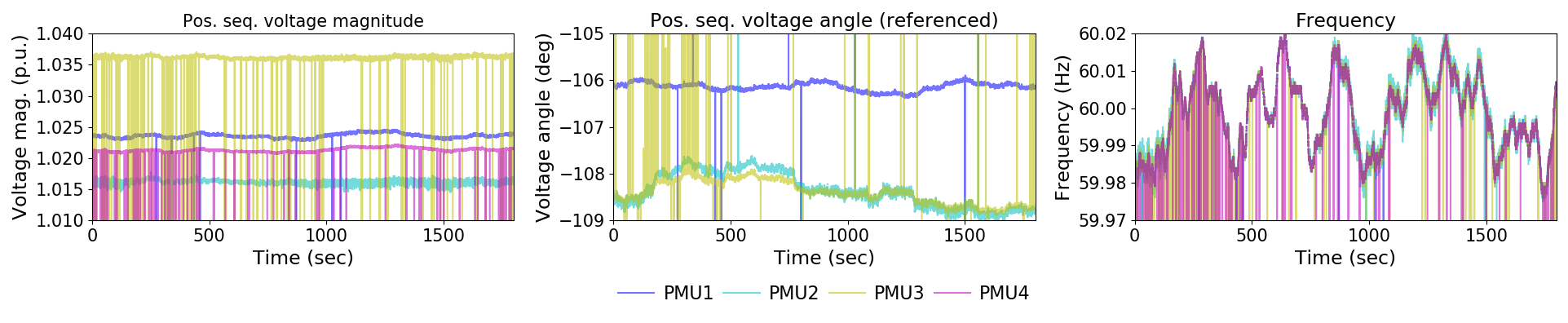}
\caption{Thirty minutes of measurements from four PMUs in an anonymized U.S. electric utility. Data is reported at 30 fps. The vertical lines show data drops.}
\label{fig:domPMU}
\includegraphics[width=\textwidth]{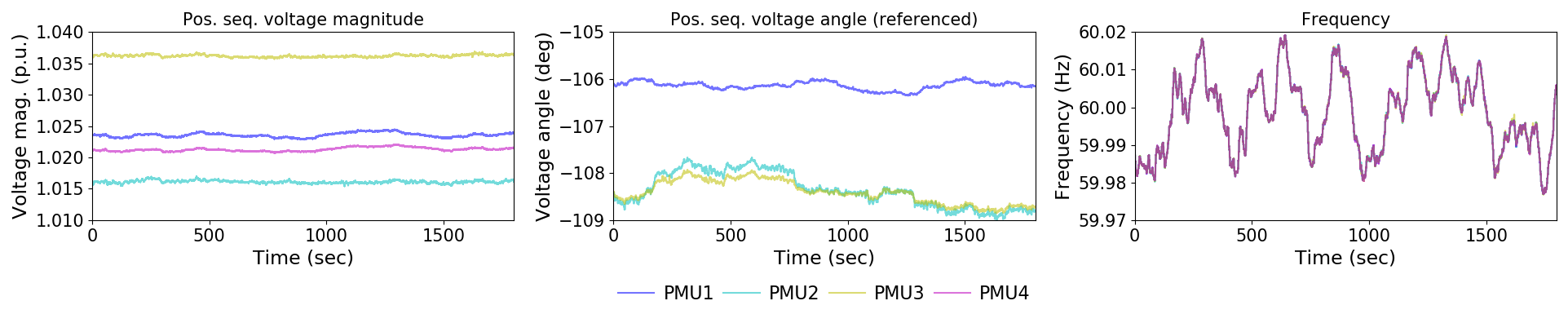}
\caption{Imputed PMU data using Page matrix}
\label{fig:cleandom}
\includegraphics[width=\textwidth]{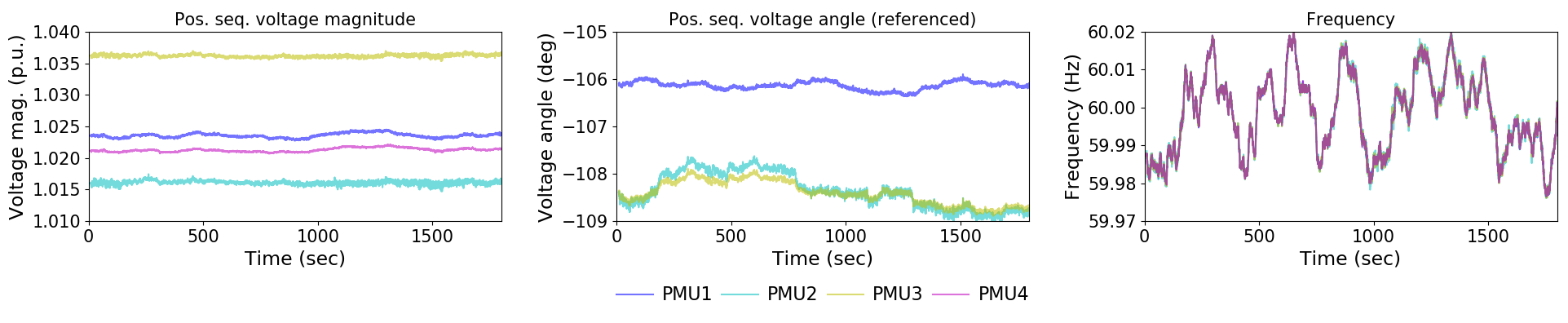}
\caption{\cmb{Imputed PMU data using Hankel matrix}}
\label{fig:cleandomHankel}
\vspace{-0.1in}
\end{figure*}

The recorded measurements were quite noisy, and had large patches of missing data. The percentage of missing entries, and maximum length of consecutive missing data segments in each channel is shown in fig. \ref{fig:dom data quality}. It can be seen that the PMU channels have data missing in the $\sim$20-40\% range. Maximum length of missing data segments is $\sim$100. The noisy and intermittent PMU measurements are shown in fig. \ref{fig:domPMU}, and the imputed measurements are shown in fig. \ref{fig:cleandom}. Since there is no way to know what the actual measurements should have been, objectively evaluating the imputation algorithm is not possible. However, visual inspection shows that the proposed algorithm is able to impute the measurements very well. Median computation time to impute 54000 measurement samples over 20 runs was 0.109 seconds.

{Figure \ref{fig:cleandomHankel} shows the imputed measurements when using the Hankel matrix method (in algorithm 2, measurements are transformed into a stacked Hankel matrix in place of a Page matrix). Visually, it appears that the Hankel method does not provide better estimates than the Page method. Moreover, time taken for imputation using the Hankel method was $\sim 1.74$ seconds, much higher than the Page method.}

\begin{figure*}
\centering
\includegraphics[width=\textwidth]{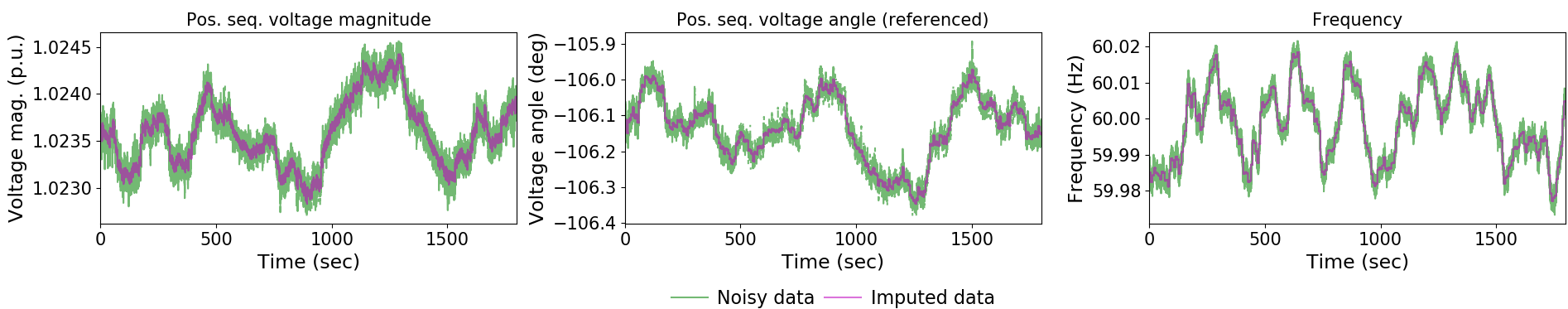}
\caption{\cmb{Imputed readings for PMU1 after additional noise is injected into its channels}}
\label{fig:domPMUnoisy}
\vspace{-0.08in}
\end{figure*}

{Further, additional $2\%$ noise was injected into the channels of PMU1 and recovered using the proposed offline imputation method using Page matrix. The noisy and imputed data is shown in figure \ref{fig:domPMUnoisy}. It can be inferred that the proposed imputation method is able to recover data from real noisy PMU archives.}

\section{Conclusion} \label{sec:4_conclusion}

In this work, a simple model-agnostic data recovery method based on low-rank matrix approximation has been proposed for improving the quality of phasor measurements with additive noise and data drop. The method is applicable for both single and multiple measurement channels; and can deal with simultaneous and consecutive data drop on all channels. Two variations  of  the  recovery  algorithm  are  shown-  a)  an  offline  block-processing method   for   imputing   past   measurements,   and   b)   an   online method for predicting future measurements. The performance of the proposed algorithms have been illustrated through extensive numeric experiments on simulated measurements on the IEEE 39-bus test system. It is seen that the proposed methodology has high accuracy, has low memory requirement and is computationally faster than other methods in literature. The performance of the algorithm is independent of the underlying system model, topology changes, and measurement noise distribution. Tests on real PMU data validate the performance of the data recovery strategy put forth in this work. The fast computation speed and ease of implementation make the algorithms developed in this work suitable for quick deployment.

\bibliographystyle{IEEEtran}
\bibliography {bib_PMU,new_ref}

\end{document}